\documentclass{optica-article}

\journal{opticajournal}

\articletype{Research Article}

\numberwithin{equation}{section}

\usepackage[utf8]{inputenc}
\usepackage[british]{babel}
\usepackage{amsmath,amsfonts}
\usepackage{cancel}
\usepackage{bm}
\usepackage{comment}
\usepackage{xcolor}
\usepackage{graphicx}
\usepackage{mathtools}
\usepackage{float}
\usepackage{enumitem}
\usepackage{soul}
\usepackage{graphics, setspace}
\usepackage{hyperref}
\usepackage{booktabs}
\usepackage{lscape}
\usepackage{rotating}
\usepackage{subcaption}

\usepackage{pgf,tikz,pgfplots}
\usetikzlibrary{arrows}
\usetikzlibrary{shapes}
\usetikzlibrary{intersections}
\usetikzlibrary{calc}
\usetikzlibrary{math}
\usetikzlibrary{decorations.markings}

\definecolor{tuewarmred}{rgb}{0.969,0.192,0.192}
\definecolor{tueprocesscyan}{rgb}{0.000,0.635,0.871}
\definecolor{tuecyan}{rgb}{0.000,0.635,0.871}
\definecolor{tuered}{rgb}{0.839,0.000,0.290}
\definecolor{tueblue}{rgb}{0.000,0.400,0.800}
\definecolor{tuedarkblue}{rgb}{0.063,0.063,0.451}
\definecolor{tueorange}{rgb}{1.000,0.604,0.000}
\definecolor{tueyellow}{rgb}{1.000,0.867,0.000}
\definecolor{tuelightgreen}{rgb}{0.518,0.824,0.000}
\definecolor{tuegreen}{rgb}{0.000,0.675,0.510}

\newcommand\numberthis{\addtocounter{equation}{1}\tag{\theequation}}

\usepackage{lineno}

\begin{document}

\title{Computing aberration coefficients for plane-symmetric reflective systems: A Lie algebraic approach}

\author{A. Barion,\authormark{1,*} M. J. H. Anthonissen,\authormark{1}
        J. H. M. ten Thije Boonkkamp,\authormark{1} and
        W. L. IJzerman\authormark{1,2}}

\address{\authormark{1}Eindhoven University of Technology, PO Box 513, 5600 MB Eindhoven, The Netherlands\\
\authormark{2}Signify, High Tech Campus 7, 5656 AE Eindhoven, The Netherlands}

\email{\authormark{*}a.barion@tue.nl} 



\begin{abstract}
We apply the Lie algebraic method to reflecting optical systems with plane-symmetric freeform mirrors. Using analytical ray-tracing equations we construct an optical map. The expansion of this map gives us the aberration coefficients in terms of initial ray coordinates. The Lie algebraic method is applied to treat aberrations up to arbitrary order. The presented method provides a systematic and rigorous approach to the derivation, treatment and composition of aberrations in plane-symmetric systems. We give the results for second- and third-order aberrations and apply them to three single-mirror examples.
\end{abstract}

\section{Introduction}
	Off-axis mirror systems provide additional degrees of freedom for the design of more compact and accurate imaging systems compared to rotationally symmetric ones. The study of their aberration behaviour is of great interest to the optics community. The mathematical characterization of aberrations has been investigated in \cite{Moore,MooreErr}. Possible approaches to derive explicit aberration expansions are given in \cite{Chang}, where only confocal arrangements are considered, or in \cite{Korsch} where the starting design is rotationally symmetric. Recently, explicit expressions for plane-symmetric reflective optical systems have been determined using a matrix formalism \cite{Caron, Caron2, Caron3}. Here, the matrix method for paraxial ray-tracing is extended to accommodate for higher degree polynomial terms and aberrations are composed by manipulating the respective matrix coefficients.
	
	In this work we will describe the Lie algebraic method needed to obtain analytical expressions for the aforementioned aberration terms of arbitrary order. Starting from a chosen ray, which we will define as the optical axis ray (OAR), we will follow its path through the system from object to image plane. At the image plane the aberration terms are given as polynomials in $(\bm{q},\bm{p})$, which are the phase-space variables of our optical system \cite{Dragt82,Wolf2004}. In this Hamiltonian formulation, the propagation and reflection maps are symplectic, i.e., volume preserving in phase-space. Our goal is to approximate these maps while preserving symplecticity. Applying these approximating maps to the initial coordinates will deliver the desired aberration expansion terms.
	
	The Lie approach provides the tools to systematically determine the approximating map for one single plane-symmetric mirror. The description of a complete optical system is then reduced to a concatenation of maps. This process is described and handled by the Lie theory. Compared to the matrix formalism in \cite{Caron, Caron2, Caron3}, the mathematical framework of the Lie method reduces the number of coefficients necessary to be stored. Additionally, the phenomenon of low-order aberrations composing into higher order contributions follows directly from the mathematical framework. This is also known as the distinction between intrinsic and extrinsic aberrations \cite{Sasian}, where low-order aberrations of individual surfaces (intrinsic) combine into higher order contributions to the complete system (extrinsic).
	
	In Section \ref{sec::analytic} we will describe the explicit maps that govern ray propagation and reflection in a plane-symmetric reflective optical system. A brief summary of the essential Lie algebraic notions is given in Section \ref{sec::LieTools}, even though we refer to \cite{DragtFinn, Wolf2004} for a more in depth description. Section \ref{sec::fundElem} contains the steps needed to construct the approximation maps and the calculations up to third-order aberrations. Three examples to validate the presented method are given in Section \ref{sec::Examples}, where both existing theoretical and computational results are reproduced.
	
	\section{Analytic Ray-Tracing}\label{sec::analytic}
\begin{figure}[!htb]
		\centering
		\includegraphics[width=0.5\textwidth]{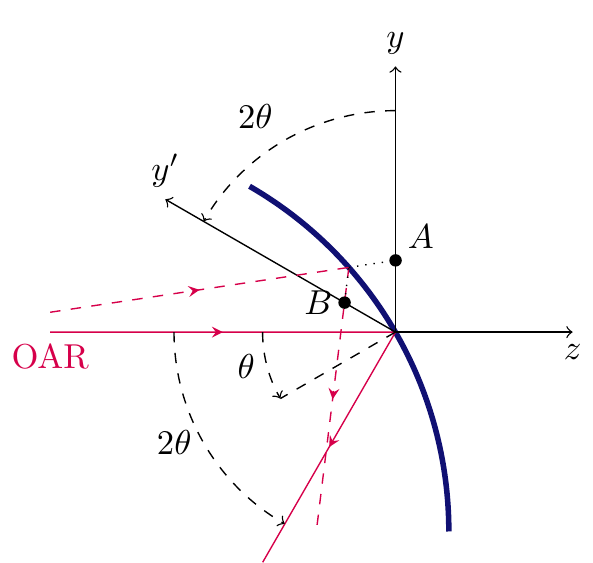}
		\caption{Point $A$ of the incoming ray in $xy$-coordinates is mapped to point $B$ of the outgoing ray in the rotated $x'y'$-coordinate system. The axes $x$ and $x'$ are perpendicular to the $yz$-plane (not shown) and the incidence angle of the OAR is equal to $\theta$.}
		\label{fig::reflectionTilt}
	\end{figure}

	In this section we discuss the mappings needed to ray-trace light rays through a reflective system composed of plane-symmetric, i.e., symmetric with respect to the $yz$-plane, optical surfaces; see Figure \ref{fig::reflectionTilt}. In order to follow a ray path from object to image plane, we describe three transformations. First, the incoming ray is propagated from the object plane to the reflecting surface and the reflected ray from the surface to the image plane. Second, we describe the reflection of the ray at a plane-symmetric mirror. Finally, rotation of the coordinate system is shown, such that the $z$-axis remains aligned with the optical axis ray (OAR) before and after reflection. This implies that the considered $z$-axis will be broken into line segments. 
	
	Once these three mappings have been described, they are concatenated to describe a single mirror, which we call the \textit{fundamental element}, according to the following five steps: $i)$ propagation from object plane to mirror; $ii)$ rotation of the optical axis and corresponding coordinate system by an angle $\theta$, equal to the incidence angle of the OAR; $iii)$ reflection of the rays; $iv)$ second rotation of coordinate system by the angle $\theta$ and $v)$ propagation from mirror to image plane.
	
	Position coordinates of an arbitrary ray before and after reflection are projected along the ray onto the two planes passing through the point of impact of the OAR and orthogonal to it; see Figure \ref{fig::reflectionTilt}. The incoming (outgoing) plane, which is orthogonal to the incoming (outgoing) OAR, will be called the incoming (outgoing) standard screen and the incoming (outgoing) position and direction coordinates will be evaluated with respect to it. The incoming standard screen is the $xy$-plane and the outgoing one is the $x'y'$-plane, where the $x$ and $x'$-axis are the same; see Figure \ref{fig::reflectionTilt}.
	
	In the remaining part of this section the three elementary maps are described independently from each other. Eventually, we concatenate them to describe a complete mirror element as previously described. Each ray is characterized by its position $\bm{q}=(q_x,q_y)$ and its direction $\bm{p}=(p_x,p_y)$ at a standard screen. As such, we use the phase-space coordinates $(\bm{q},\bm{p})$ as our ray coordinates, cf.  \cite{Barion:22}. Note that the coordinates of the OAR are at the origin of phase-space both before and after reflection, i.e., the OAR will have coordinates $\bm{q}=\bm{0}=\bm{q}'$ and $\bm{p}=\bm{0}=\bm{p}'$. In the descriptions to follow phase-space coordinates $(\bm{q},\bm{p})$ are mapped to primed coordinates $(\bm{q}',\bm{p}')$ by the respective mappings.
	
	\subsection{Propagation}
	We introduce the Hamiltonian $H(\bm{p})$ governing free propagation of light in a medium of constant refractive index $n$ \cite{DragtFoundations86,Wolf2004,Barion:22}, i.e.,
	\begin{equation}
		\label{eq::Hamiltonian}
		H(\bm{p})=-\sigma\sqrt{n^2-\vert\bm{p}\vert^2}=-\sigma p_z,
	\end{equation}
	where $\bm{p}=(p_x,p_y)$ and $p_z$ are the direction momenta -- direction cosines times the refractive index $n$ -- along the respective axes. The variable $\sigma=\pm 1$, is positive for forward travelling rays and negative for backwards propagating rays. Since we are only considering reflections, the refractive index of our medium (air/vacuum) $n=1$. The distance measured along the optical axis, which coincides with the $z$-axis, serves as evolution parameter of the Hamiltonian system related to Eq.~\eqref{eq::Hamiltonian}:
	\begin{equation}
		\label{eq::HamSys}
		\dot{\bm{q}}=\frac{\partial H}{\partial\bm{p}}=-\frac{\bm{p}}{H},\qquad\dot{\bm{p}}=-\frac{\partial H}{\partial\bm{q}}=\bm{0}.
	\end{equation} 
	The solution to the Hamiltonian system Eq.~\eqref{eq::HamSys} with initial conditions $(\bm{q},\bm{p})$, after propagating a distance $d$ along the optical axis ray, reads:
	\begin{equation}
		\label{eq::hamSolution}
			\bm{q}'=\bm{q}-d\frac{\bm{p}}{H(\bm{p})},\quad
			\bm{p}'=\bm{p}.
	\end{equation}
	
	\subsection{Reflection}
	Next, we consider the law of reflection in vector form regardless of the coordinate system \cite{Welford1986}
	\begin{equation}
		\label{eq::lawOfReflection}
		\hat{\bm{k}}_\mathrm{r}=\hat{\bm{k}}_\mathrm{i}-2(\hat{\bm{k}}_\mathrm{i}\cdot\hat{\bm{n}})\hat{\bm{n}},
	\end{equation}
	where $\hat{\bm{k}}_\mathrm{r}$ is the unit direction vector of the reflected ray, $\hat{\bm{k}}_\mathrm{i}$ the unit direction vector of the incoming ray and $\hat{\bm{n}}$ the unit outer normal of the reflector at the impact point. Here, $\hat{}$ (hat) indicates that the vector has length one and with the term `outer' we mean opposite to the incoming ray direction, i.e., $\hat{\bm{k}}_\mathrm{i}\cdot\hat{\bm{n}}<0$. Let the reflector be described by $z=\zeta(\bm{q})$, then the outer normal $\hat{\bm{n}}$ of the surface at point $(\bm{q},\zeta(\bm{q}))$ reads
	\begin{equation}
		\label{eq::usedNormal}
		\hat{\bm{n}}=\frac{(\nabla\zeta(\bm{q}),-1)}{\sqrt{1+\vert\nabla\zeta(\bm{q})\vert^2}}.
	\end{equation}
	The incoming and outgoing ray directions are $\hat{\bm{k}}_\mathrm{i}=(\bm{p},p_z)/n$ and $\hat{\bm{k}}_r=(\bm{p}',p_z')/n$, respectively. The vector $\hat{\bm{k}}_\mathrm{r}$ is calculated inserting Eq.~\eqref{eq::usedNormal} in Eq.~\eqref{eq::lawOfReflection}. This way we get for the reflected momenta $(\bm{p}',p_z')$:
	\begin{subequations}
		\label{eq::reflectedMomentum}
		\begin{equation}
			\bm{p}'=\bm{p}-2\frac{\nabla\zeta(\bar{\bm{q}})}{1+\vert\nabla\zeta(\bar{\bm{q}})\vert^2}(\bm{p}\cdot\nabla\zeta(\bar{\bm{q}})-p_z),
		\end{equation}
		\begin{equation}
			p_z'=p_z+\frac{2}{1+\vert\nabla\zeta(\bar{\bm{q}})\vert^2}(\bm{p}\cdot\nabla\zeta(\bar{\bm{q}})-p_z),
		\end{equation}
	\end{subequations} 
	where $(\bar{\bm{q}},\zeta(\bar{\bm{q}}))$ is the intersection point of the incoming ray and the reflector. The intersection point $\bar{\bm{q}}$ is related to the screen coordinate before reflection $\bm{q}$ and the one after reflection $\bm{q}'$ by \cite{Wolf2004,DragtFoundations86, Barion:22}
	\begin{equation}
		\label{eq::qBar}
		\bar{\bm{q}}=\bm{q}+\zeta(\bar{\bm{q}})\frac{\bm{p}}{p_z},\quad
		\bm{q}'=\bar{\bm{q}}-\zeta(\bar{\bm{q}})\frac{\bm{p}'}{p_z'}.
	\end{equation}
	Eq.~\eqref{eq::qBar} gives an implicit relation for $\bar{\bm{q}}$ which needs to be solved iteratively; see \cite{Barion:22,SaadWolf1986,DragtFoundations86}. The Eqs.~\eqref{eq::qBar} are again solution to the Hamiltonian system Eq.~\eqref{eq::HamSys}, but now propagating a distance $d=\zeta(\bar{\bm{q}})$.
	
	\subsection{Rotation of the Standard Screen}
\begin{figure}[!htb]
		\centering
		\includegraphics[width=0.4\textwidth]{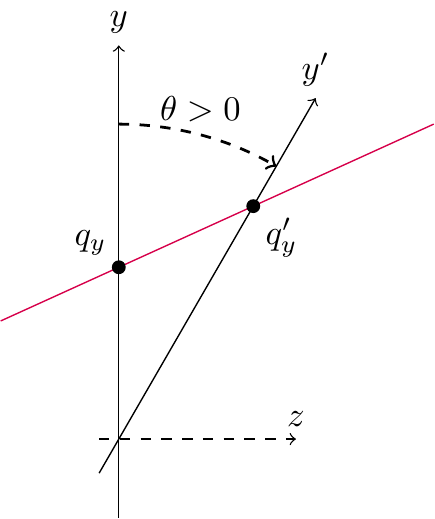}
		\caption{Upon rotation of axes we map the position coordinate $q_y$ to the position coordinate $q_y'$ relating to the same ray (red).}
		\label{fig::rotPosition}
	\end{figure}

	After propagation and reflection, we discuss the necessary steps to rotate our coordinates according to the OAR. We describe an arbitrary rotation by an angle $\theta$ that rotates our standard screen, see Figure~\ref{fig::rotPosition}. For a single reflector two rotations of angle $\theta$ are used, where eventually $\theta$ is the incidence angle of the OAR. The first rotation brings the $z$-axis of the incoming coordinate system from being aligned with the OAR to being aligned with the surface normal at the point of intersection of the OAR. The surface equation $z=\zeta(\bm{q})$ is defined in this coordinate system aligned with its normal and therefore has zero gradient at the origin, i.e., $\nabla\zeta(\bm{0})=\bm{0}$, which is the point of impact of the OAR. We then apply the reflection mapping and subsequently rotate to the outgoing coordinate system aligned with the reflected OAR, see Figure~\ref{fig::rotationCoordSys}.
	
	Let us define positive rotations when the $y$-axis is rotated towards the $z$-axis (clock-wise). In the starting coordinate system the $z$-axis is aligned with the incoming OAR and as such we can call it the incoming coordinate system. We consider surfaces with plane-symmetry with respect to the $yz$-plane and as such the rotations are around the $x$-axis. 
	
	The rotation mapping of the screen around the $x$-axis for the phase-space coordinates can be found as a Lie transformation \cite{Wolf2004}. Here we present an equivalent derivation. The momentum coordinates $p_x,p_y,p_z$ are rotated into the coordinates $p_x',p_y',p_z'$ according to the well-known rotation matrix
	\begin{equation}
		\label{eq::rotMomentum}
			\begin{pmatrix}
				p_x' \\ p_y' \\ p_z'
			\end{pmatrix}	=\begin{pmatrix}
				1 & 0 & 0\\ 0 & \cos\theta & \sin\theta\\ 0 & -\sin\theta & \cos\theta
			\end{pmatrix}\begin{pmatrix}
			p_x \\ p_y \\ p_z
		\end{pmatrix}.
	\end{equation}
	The expressions for the position coordinates are more complex. In fact, recall that we map the intersection points of the light rays with the (rotated) standard screens; see Figure \ref{fig::rotPosition}.
	As such, let us first fix the parametrization of the ray and the normal equation of the tilted surface. The path of the ray can be parametrized by
	\begin{equation}
		\binom{\bm{q}}{0}+\lambda\binom{\bm{p}}{p_z},\quad\lambda\in\mathbb{R}.
		\label{eq::rayEq}
	\end{equation}
	After the first rotation, the equation of the rotated standard screen, with normal $(0,\sin\theta,-\cos\theta)$ and passing through $(0,0,0)$, reads
	\begin{equation}
		y\sin\theta-z\cos\theta=0.
		\label{eq::tiltedScreenEq}
	\end{equation} 
	By substituting the parametrization \eqref{eq::rayEq} into the rotated standard screen equation \eqref{eq::tiltedScreenEq} we can solve for $\lambda$ to get the point of intersection. We get
	\begin{equation}
		\lambda=\frac{q_y\sin\theta}{p_z\cos\theta-p_y\sin\theta}.
		\label{eq::lambda}
	\end{equation}
	Substituting the value in Eq.~\eqref{eq::lambda} in the parametrization \eqref{eq::rayEq} gives us the coordinates of the point of intersection of the considered ray and the tilted screen. The last step is to derive the position coordinates with respect to the rotated coordinate system, which corresponds to dividing the $y$-coordinate by $\cos\theta$. The map for positive rotation around the $x$-axis of the position coordinates $\bm{q}$ by an angle $\theta$ reads:
	\begin{subequations}
		\label{eq::rotPosition}
		\begin{equation}
			q_x'=\frac{q_x\,p_z\cos\theta-(q_x\,p_y-q_y\,p_x)\sin\theta}{p_z\cos\theta-p_y\sin\theta},\\
		\end{equation}
		\begin{equation}
			q_y'=\frac{q_y\,p_z}{p_z\cos\theta-p_y\sin\theta}.
		\end{equation}
	\end{subequations}
 	In Eq. \eqref{eq::rotPosition} the condition $p_z\cos\theta-p_y\sin\theta=0$ implies that the considered ray is parallel to the rotated plane and as such will not intersect with the plane.

	\subsection{The Fundamental Map}\label{sec::fundElemLast}
	With the transformations described in Eqs.~\eqref{eq::reflectedMomentum}-\eqref{eq::rotMomentum} and \eqref{eq::rotPosition} we can rotate the coordinate system for the $z$-axis to be aligned with the surface normal, reflect the incoming rays and rotate the system again to align the $z$-axis with the outgoing OAR. If propagation before and after the surface are added to this map, we will call it the \textit{fundamental map}. This composition of transformations can be expanded up to the desired order in terms of the phase-space coordinates $(\bm{q},\bm{p})$.

\begin{figure}[!htb]
		\centering
		\includegraphics[width=0.5\textwidth]{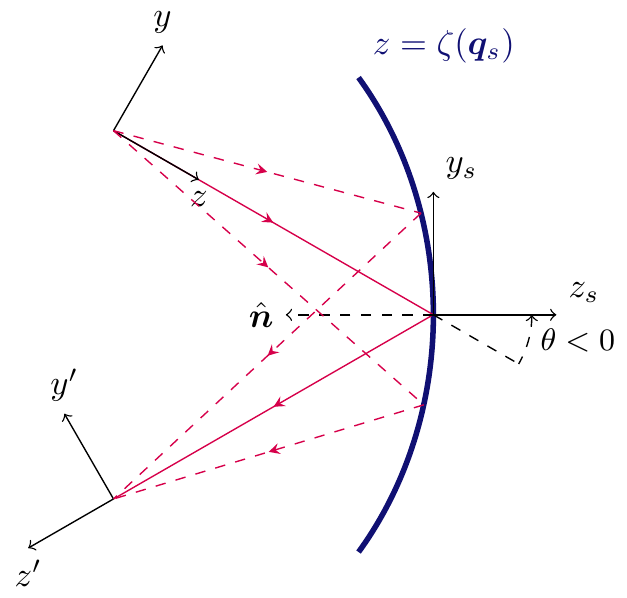}
		\caption{Definition of incoming $xyz$ and outgoing $x'y'z'$-coordinate systems. The auxiliary system denoted by the index $s$ is where the surface equation is defined.}
		\label{fig::rotationCoordSys}
	\end{figure}

	After reflection, $\sigma$ in the Hamiltonian described in Eq.~\eqref{eq::Hamiltonian} changes sign. An intuitive way to understand this is to recall Eq.~\eqref{eq::Hamiltonian} with $\sigma=1$ where $H=-p_z$. By the condition $\hat{\bm{k}}_\mathrm{i}\cdot\hat{\bm{n}}<0$ that we imposed at reflection, the reflected OAR travels in the same $z$-direction as the surface normal, which is opposite to the one of the incoming OAR. This would lead to negative propagation distances. Since we prefer to consider forward moving rays, we opt for dealing with a left-handed coordinate system and align the $z'$-axis in Figure \ref{fig::rotationCoordSys} with the direction of the reflected OAR after the second rotation. It can be verified that this change does not influence the form of our rotation map and the reflection map remains also unchanged. The only important caveats are that the reflective surface must always be described in the coordinate system of the incoming OAR and that angles are positive when the $y$-axis rotates towards the $z$-axis.
	
	With the mappings described in Eqs.~\eqref{eq::reflectedMomentum}-\eqref{eq::rotMomentum} and \eqref{eq::rotPosition} we can concatenate them into a reflection plus rotation mapping. We define this composition of transformations by $\mathcal{S}(\theta)$. Let $\mathcal{R}(\theta)$ denote the rotation mapping by an angle of $\theta$ and $\mathcal{T}$ the reflection mapping. Then, we can concisely describe the map $\mathcal{S}(\theta)$ as
	\begin{equation}
		\label{eq::reflWithRotMap}
		\mathcal{S}(\theta)=\mathcal{R}(\theta)\,\mathcal{T}\,\mathcal{R}(\theta).
	\end{equation}	
	In Figure \ref{fig::reflectionTilt}, we have that $\mathcal{S}(\theta)$ maps $A$ to $B$. This definition of $\mathcal{S}(\theta)$ is necessary to apply the Lie algebraic method. Note that the surface equation is given with respect to the coordinate system denoted by the index $s$ in Figure~\ref{fig::rotationCoordSys}.
	
	To conclude, concatenating $\mathcal{S}(\theta)$ with propagation in object and image-space, $\mathcal{P}_{\mathrm{ob}}$ and $\mathcal{P}_{\mathrm{im}}$ respectively, constitutes the fundamental map $\mathcal{M}$ necessary for our description of the optical system
	\begin{equation}
		\label{eq::fundMap}
		\mathcal{M}=\mathcal{P}_{\mathrm{im}}\mathcal{S}(\theta)\mathcal{P}_{\mathrm{ob}}.
	\end{equation}

	\section{Lie Algebraic Tools}\label{sec::LieTools}
	With the help of the Lie algebraic method it is possible to construct operators that reproduce the actions of propagation, reflection and rotation. These operators enable us to derive closed form expressions for the aberration components of an arbitrary optical system. A more detailed description of the Lie algebraic tools used in this work can be found in \cite{Barion:22,Wolf2004,DragtFoundations86}. Here, we briefly introduce the main concepts.
	
	The space of functions on phase-space becomes a Lie algebra when endowed with the Poisson bracket $[\cdot,\cdot]$. The Poisson bracket of two functions $f(\bm{q},\bm{p}) ,g(\bm{q},\bm{p})$ is defined as
	\begin{equation}
		\label{eq::poissonBracket}
		[f,g]=\frac{\partial f}{\partial \bm{q}}\boldsymbol{\cdot}\frac{\partial g}{\partial \bm{p}}-\frac{\partial f}{\partial \bm{p}}\boldsymbol{\cdot}\frac{\partial g}{\partial \bm{q}}.
	\end{equation}
	Accordingly, we can associate with each $f$ a Lie operator $[f,\cdot\,]$ that acts on a second function $g$ by taking the Poisson bracket of the two. For example, $[q_1,\cdot\,]=\partial\cdot/\partial p_1$ and for vectors we have $[\bm{q},\cdot\,]=\partial\cdot/\partial \bm{p}$.
	
	Using the Poisson bracket, we can associate to each function $f$ on phase-space a mapping $\exp([f,\cdot\,])$, called a Lie transformation, defined as
	\begin{equation}
		\label{eq::LieTransformation}
		\exp([f,\cdot\,])=\sum_{k=0}^\infty \frac{[f,\cdot\,]^k}{k!},
	\end{equation}
	where $[f,\cdot\,]^0=I$ and $[f,\cdot\,]^k=[f,[f,\cdot\,]^{k-1}]$ for $k>1$. Suppose $f$ is only dependent on $\bm{q}$, i.e. $f=f(\bm{q})$, then 
	\begin{equation}
		\label{eq::LieTransExample}
		\exp([f(\bm{q}),\cdot\,])\bm{q}=\bm{q}\quad\text{and}\quad\exp([f(\bm{q}),\cdot\,])\bm{p}=\bm{p}+\frac{\partial f}{\partial\bm{q}}.
	\end{equation} 
	In Eq.~\eqref{eq::LieTransExample} the infinite series is truncated after the first two terms as any subsequent one is equal to zero. Note that Lie transformation are applied component-wise to vectors.
	
	A map $(\bm{q},\bm{p})\mapsto (\bm{q}'(\bm{q},\bm{p})$, $\bm{p}'(\bm{q},\bm{p}))$ is said to be a symplectic transformation, if it satisfies \cite{Wolf2004,DragtFinn}:
	\begin{align}
		[q'_i,q'_j]&=[q_i,q_j]=0,\nonumber\\
		[p'_i,p'_j]&=[p_i,p_j]=0,\label{eq::canonicalTransformation}\\
		[q'_i,p'_j]&=[q_i,p_j]=\delta_{ij},\nonumber
	\end{align}
	where $\delta_{ij}$ is the Kronecker delta. Symplectic transformations preserve volumes in phase-space. In fact, light propagation, reflection and rotation are all symplectic maps.
	
	It can be proven that a mapping defined as in Eq.~\eqref{eq::LieTransformation} is symplectic \cite{DragtFinn}. Conversely, symplectic mappings $\mathcal{M}$ that map the origin to itself, i.e., $\mathcal{M}(\bm{0})=\bm{0}$, can be represented as an infinite concatenation of Lie transformations of the form
		\begin{equation}
		\label{eq::thrm2}
		\mathcal{M}=\exp([g_2,\cdot\,])\exp([g_3,\cdot\,])\cdots,
	\end{equation}
	where the generators $g_2,g_3,$ etc.\ are homogeneous polynomials in the variables $(\bm{q},\bm{p})$ of degree $2,3,$ etc.~\cite{DragtFinn}. Here, we omit the concatenation symbol $\circ$, as it is clear from the context that we are concatenating operators. Recall that a homogeneous polynomial $g$ of degree $m$, as in Eq. \eqref{eq::thrm2}, has the following property
	\begin{equation}
		g(\lambda\bm{q},\lambda\bm{p})=\lambda^mg(\bm{q},\bm{p})\quad\forall\lambda\in\mathbb{R}.
	\end{equation}

	The maps for ray propagation and reflection plus rotation are symplectic and map the origin onto itself \cite{Wolf2004,DragtFoundations86,Barion:22}. It is therefore possible to represent them as an infinite, or approximate them by a truncated, concatenation of Lie transformations according to the result in Eq.~\eqref{eq::thrm2} and then rearrange the Lie transformations using additional Lie tools given in Appendix \ref{sec::AddLieTools}, cf. Eq.~\eqref{eq::BCH} and Eq.~\eqref{eq::thrm3}. 
	
	Our aim is to approximate the fundamental map $\mathcal{M}$ in Eq.~\eqref{eq::fundMap} of a reflector by means of a truncated concatenation of Lie transformations in ascending order, similarly to the structure of Eq.~\eqref{eq::thrm2}. This enables us to clearly distinguish which parts of the map influence which order of aberrations. In fact, generators of order $k$ are directly related to the transverse ray aberrations of order $k-1$ \cite{Barion:22}. Concatenating multiple fundamental maps representing the different mirrors in our system and disregarding terms that lead to higher order aberrations leads to a map describing the complete optical system -- up to the desired order of accuracy in terms of initial phase-space coordinates.
	
	\section{The Fundamental Element}\label{sec::fundElem}
	Free propagation, reflection and rotation are symplectic maps and their combined actions map the origin of phase-space, i.e., the OAR, to itself. As such, it is possible to represent the combined actions of reflection and rotation $\mathcal{S}(\theta)$, see Eq.~\eqref{eq::reflWithRotMap}, in the form of Eq.~\eqref{eq::thrm2}. The polynomials necessary for this representation in terms of Lie transformations are called the \textit{generators} of the map. It is important to consider the complete reflection with rotation map $\mathcal{S}(\theta)$ because this ensures that the origin of phase-space, i.e., our OAR, is mapped onto itself. Hence, we have that $(\mathcal{S}(\theta))(\bm{0})=\bm{0}$ and we can apply the results in Eq.~\eqref{eq::thrm2}. Rotation alone does not map the origin of phase-space onto itself.
	
	We subsequently concatenate $\mathcal{S}(\theta)$ with the maps of object and image-space propagation to derive the description of the \textit{fundamental element} of the optical system. The fundamental element represents the physical counterpart of the fundamental map described at the end of Section \ref{sec::analytic}. This fundamental element is the building block of any arbitrary reflecting optical system with plane-symmetry with respect to the $yz$-plane. We restrict our analysis to aberrations of order three and therefore only polynomials up to degree four in Eq.~\eqref{eq::thrm2} are of relevance; see \cite{Barion:22,DragtFoundations86,Wolf2004}.
	
 	The generators of free propagation for light rays in a medium of refractive index $n=1$ are, up to degree four \cite{Barion:22},
 	\begin{equation}
 		\label{eq::HamExpansion}
 		h_2(\bm{p})=\frac{1}{2}\vert\bm{p}\vert^2,\quad
 		h_4(\bm{p})=\frac{1}{8}\vert\bm{p}\vert^4.
 	\end{equation}
 	This means, that if we want to propagate our physical system with initial condition $(\bm{q},\bm{p})$ over a distance $d$ along the optical axis, then the expression
 	\begin{equation}
 		\label{eq::thirdOrderProp}
 		\binom{\bm{q}'}{\bm{p}'}=\exp(-d[h_2,\cdot\,])\exp(-d[h_4,\cdot\,])\binom{\bm{q}}{\bm{p}},
 	\end{equation}
 	is equal, up to third-order terms, to the solution given in Eq.~\eqref{eq::hamSolution} \cite{Barion:22,DragtFoundations86,Wolf2004}. The result of Eq.~\eqref{eq::thirdOrderProp} is therefore sufficiently accurate to investigate third-order aberrations. Note that the polynomials in Eq.~\eqref{eq::HamExpansion} are simply the first two terms in the Taylor expansion of the Hamiltonian defined in Eq.~\eqref{eq::Hamiltonian}.
 	
 	The mirror equation is given in the coordinate system with its $z$-axis aligned with the surface's normal and is of the form
	\begin{equation}
		\label{eq::surfaceEq}
		z=\zeta(\bm{q})=\sum_{\substack{2\leq m+n\leq 4 \\ m \text{ even}}}c_{mn}q_x^mq_y^n.
	\end{equation}
	We consider surface terms up to fourth order, since higher order terms do not influence third-order aberrations. 
	
	The reflection and rotation mapping $\mathcal{S}(\theta)$ maps $\bm{q},\bm{p}$ to $\bm{q}',\bm{p}'$. First, the rotation by the angle $\theta$ is applied to the incoming ray coordinates, cf. Eqs.~\eqref{eq::rotMomentum},\eqref{eq::rotPosition}. Secondly, reflection acts on these already rotated coordinates, cf. Eqs.~\eqref{eq::reflectedMomentum},\eqref{eq::qBar}. Lastly, a second rotation by $\theta$ maps these coordinates into the final reflected coordinate system. All these transformations -- and their concatenation -- can be expanded in terms of $(\bm{q},\bm{p})$ with the aid of computer algebra software, e.g., Mathematica. The first order expansion of $\mathcal{S}(\theta)$ reads:
	\begin{equation}
		\label{eq::firstOrder}
		\begin{aligned}
			q'_x&=q_x,\\
			q'_y&=q_y,\\
			p'_x&=p_x + 4  \,c_{2 0}\cos(\theta) \,q_x,\\
			p'_y&=p_y + 4  \,c_{0 2}\sec(\theta)\,q_y.
		\end{aligned}
	\end{equation}
	Here, the coefficients $c_{20},c_{02}$, cf. Eq.~\eqref{eq::surfaceEq}, can be related to the radii of curvature of the mirror surface. The polynomial $g_2$ associated with the Lie transformation that generates the linear map in Eq.~\eqref{eq::firstOrder} is only dependent on $\bm{q}$:
	\begin{equation}
		\label{eq::g2}
		g_2(\bm{q})=2\,c_{20}\cos(\theta)\,q_x^2+2\,c_{02}\sec(\theta)\,q_y^2.
	\end{equation}
	The Lie transformation generated by Eq.~\eqref{eq::g2} reads
	\begin{equation}
		\label{eq::Lieg2}
		\binom{\bm{q}'}{\bm{p}'}=\exp([g_2(\bm{q}),\cdot\,])\binom{\bm{q}}{\bm{p}}=\binom{\bm{q}}{\bm{p}+\dfrac{\partial g_2(\bm{q})}{\partial \bm{q}}}.
	\end{equation} 
	One can verify that the expression in Eq.~\eqref{eq::Lieg2} is the same as the one in Eq.~\eqref{eq::firstOrder}. Note that, if $g_2$ would also depend on $\bm{p}$, it would generate contributions to the $\bm{q}$-coordinates, which is undesired; cf. Eq.~\eqref{eq::firstOrder}.
	
	To initiate a more systematic approach, we define the generators $g_m$ in a more general way:
	\begin{equation}
		\label{eq::polyGeneral}
		g_m(\bm{q},\bm{p})=\sum_{i+j+k+l=m} a_{ijkl}\,q_x^i\,q_y^j\,p_x^k\,p_y^l,\quad i,j,k,l\in\mathbb{N},\quad i+k\text{ even}
	\end{equation}
	where the condition $i+k$ even stems from the symmetry of the optical system itself. In this notation the functions $g_2,g_3,g_4$ are defined by their coefficients. It is our goal to determine these coefficients such that 
	\begin{equation}
		\mathcal{S}(\theta)\overset{(3)}{=}\exp([g_2,\cdot\,])\exp([g_3,\cdot\,])\exp([g_4,\cdot\,]).
		\label{eq::Sapprox}
	\end{equation}
	The notation $\overset{(3)}{=}$ symbolizes that the truncated concatenation of Lie transformations on the right-hand side (RHS) of Eq.~\eqref{eq::Sapprox} produces the same expressions as the map $\mathcal{S}(\theta)$ up to third-order terms in phase-space coordinates. 
	
	To derive these coefficients, one has to expand the mapping $\mathcal{S}(\theta)$ up to the order of interest, i.e., 3 in our case. Subsequently, we consider the general form of the generators as given in Eq.~\eqref{eq::polyGeneral} and compute the action of the concatenation of Lie transformations in Eq.~\eqref{eq::Sapprox} on the phase-space variables. Since the coefficients of the generator $g_k$ are fully determined by their contributions to the aberrations of order $k-1$, we can determine the coefficients of the generators in increasing order; see the method described in \cite{Barion:22}. The non-zero coefficients of the generators of the reflection plus rotation mapping $\mathcal{S}(\theta)$ are listed in Table \ref{tbl::reflRotCoeffs} in the form described by Eq.~\eqref{eq::polyGeneral}.
	\footnotesize
	\begin{table}
		\centering
		\begin{tabular}{l p{11cm}}
			\toprule
			Coefficients & Values\\[0.5ex]
			\midrule
			 $a_{0  2  0  0}$ & $2  \,c_{0  2} \sec(\theta)$   \\
			 $a_{2  0  0  0}$ & $2 \,c_{2  0} \cos(\theta)   $  \\
			 \midrule
			 $a_{0  2  0  1}$ & $2  \,c_{0  2} \sec(\theta) \tan(\theta)$  \\
			 $a_{0  3  0  0}$ & $2 \sec(\theta)^2 ( c_{0  3} - 2  \,c_{0  2}^2 \tan(\theta))$\\
			 $a_{1  1  1  0}$ & $4  \,c_{2  0} \sin(\theta)$ \\  
			 $a_{2  0  0  1}$ & $-2  \,c_{2  0} \sin(\theta)$  \\
			 $a_{2  1  0  0}$ & $2 ( c_{2  1} - 4 \cos(\theta)  \,c_{2  0}^2 \sin(\theta) + 
			2  \,c_{0  2}  \,c_{2  0} \tan(\theta))$   \\
			\midrule
			 $a_{0  2  0  2}$ & $-\frac{1}{2} (-1 + 3 \cos(2 \theta))  \,c_{0  2} \sec(\theta)^3$\\
			 $a_{0  2  2  0}$ & $- c_{0  2} \sec(\theta) + 2  \,c_{2  0} \sin(\theta) \tan(\theta)$ \\
			 $a_{0  3  0  1}$ & $
			2 \sec(\theta)^4 (- \,c_{0  2}^2 + 3 \cos(2 \theta)  \,c_{0  2}^2 + 
			\,c_{0  3} \sin(2 \theta))  $\\
			 $a_{0  4  0 0}$ & $-\sec(\theta)^5 ( \,c_{0  2}^3 + 
			\cos(2 \theta) (7  \,c_{0  2}^3 -  c_{0  4}) -  c_{0  4} + 
			4  \,c_{0  2}  \,c_{0  3} \sin(2 \theta)) $  \\
			 $a_{1  2  1  0}$ & $4  \,c_{0  2}  \,c_{2  0} - 2  \,c_{2  0}^2 + 
			2 \cos(3 \theta)  \,c_{2  0}^2 \sec(\theta) + 4  \,c_{2  1} \tan(\theta)  $	\\
			 $a_{2  0  0  2}$ & $-\cos(\theta)  \,c_{2  0}$\\  
			 $a_{2  0  2  0}$ & $-\cos(\theta)  \,c_{2  0}$	\\
			 $a_{2  1  0  1}$ & $4  \,c_{0  2}  \,c_{2  0}$\\
			 $a_{2  2  0  0}$ & $2 (-4 \cos(\theta)  \,c_{0  2}  \,c_{2  0}^2 + 
			8  \,c_{0  2}^2  \,c_{2  0} \sec(\theta)^3 - 4  \,c_{2  0}  \,c_{2  1} \sin(\theta) + 
			\sec(\theta) (-12  \,c_{0  2}^2  \,c_{2  0} +  c_{2  2} - 	6  \,c_{0  2}^2  \,c_{2  0} \tan(\theta)^2))  $\\
			 $a_{3  0  1  0}$ & $4 \cos(\theta)^2  \,c_{2  0}^2$  \\
			 $a_{4  0  0  0}$ & $-8 \cos(\theta)^3  \,c_{2  0}^3 + 2 \cos(\theta)  \,c_{4  0} - 
			2  \,c_{0  2}  \,c_{2  0}^2 \sin(\theta) \tan(\theta)$\\
			\bottomrule
		\end{tabular}
	\caption{Coefficients of $g_2,g_3,g_4$.}
	\label{tbl::reflRotCoeffs}
	\end{table}
	\normalsize
	
	We proceed to combine reflection and rotation with propagation before and after the surface. The mapping $\mathcal{M}$ will describe the action of a fundamental element on the rays from object plane coordinates $(\bm{q},\bm{p})$ to image plane coordinates $(\bm{q}',\bm{p}')$.  The map $\mathcal{M}$ is, up to fourth degree generators, composed as follows:
	\begin{multline}
		\mathcal{M}\overset{(3)}{=}\underbrace{\exp\left(-s_\mathrm{ob}[h_2,\cdot\,]\right)\exp\left(-s_\mathrm{ob}[h_4,\cdot\,]\right)}_{\overset{(3)}{=}\text{propagation from object plane}}\underbrace{\exp([g_2,\cdot\,])\exp([g_3,\cdot\,])\exp([g_4,\cdot\,])}_{\overset{(3)}{=}\mathcal{S}(\theta)}\\
		\underbrace{\exp\left(-s_\mathrm{im}[h_2,\cdot\,]\right)\exp\left(-s_\mathrm{im}[h_4,\cdot\,]\right)}_{\overset{(3)}{=}\text{propagation to image plane}}.
		\label{eq::fundElement}
	\end{multline}
	Here, $s_\mathrm{ob},s_\mathrm{im}$ are the object and image distances measured along the OAR in the sagittal plane. Although it might appear counter-intuitive, due to their unique properties, the order of Lie transformations is left-to-right like the order of transformations undergone by the ray \cite{Wolf2004}. The object and image distances for the sagittal and tangential planes satisfy the Coddington equations \cite{Braat2019}:
	\begin{subequations}
		\label{eq::coddington}
		\begin{align}
			\text{sagittal plane}{:}\quad&\frac{1}{s_\mathrm{ob}}+\frac{1}{s_\mathrm{im}}=-4\,c_{20}\cos(\theta),\\ 
			\text{tangential plane}{:}\quad&\frac{1}{t_\mathrm{ob}}+\frac{1}{t_\mathrm{im}}=-4\,c_{02}\sec(\theta).
		\end{align}
	\end{subequations}
	
	We want to reorder and combine the Lie transformations of Eq.~\eqref{eq::fundElement} into three Lie transformations generated by the functions $\tau_2,\tau_3,\tau_4$ such that 
	\begin{equation}
		\label{eq::fundElementMod}
		\mathcal{M}\overset{(3)}{=}\exp([\tau_2,\cdot\,])\exp([\tau_3,\cdot\,])\exp([\tau_4,\cdot\,]).
	\end{equation}
	This allows us to separate the linear part of the mapping, generated by $\tau_2$, from the higher order parts generated by $\tau_3,\tau_4$ that induce aberrations. Again, equality up to third-order expansions is sufficient for our current work since we are investigating aberrations up to this same order. The functions $\tau_2,\tau_3,\tau_4$ describe the action of a fundamental element up to the expansion order $3$. To derive the functions $\tau_2,\tau_3,\tau_4$ it is necessary to manipulate the mapping in Eq.~\eqref{eq::fundElement} such that the generators are combined and reordered in ascending order. The procedure has been shown in \cite{Barion:22} and a short example can be found in Appendix \ref{sec::AddLieTools}. The main tools necessary for these calculations are the Baker-Campbell-Hausdorff (BCH) formula \eqref{eq::BCH} and the identity given in Eq.~\eqref{eq::thrm3}. 
	
	The Lie transformation generated by the second degree polynomial $\tau_2$ is more conveniently represented in its matrix form $M_G$ as it is the matrix multiplication of its three components, i.e., object-space propagation, reflection with rotation and image-space propagation. We call this the Gaussian part of the mapping $\mathcal{M}_G=\exp([\tau_2,\cdot\,])$ and the associated $M_G$ reads
	\small
	\begin{equation}
		M_G=\begin{pmatrix}
			1 & 0 & s_\mathrm{im} & 0\\
			0 & 1 & 0 & s_\mathrm{im}\\
			0 & 0 & 1 & 0\\
			0 & 0 & 0 & 1			
		\end{pmatrix}
		\begin{pmatrix}
			1 & 0 & 0 & 0\\
			0 & 1 & 0 & 0\\
			4\,c_{20}\cos(\theta) & 0 & 1 & 0\\
			0 & 4\,c_{02}\sec(\theta) & 0 & 1
		\end{pmatrix}
		\begin{pmatrix}
			1 & 0 & s_\mathrm{ob} & 0\\
			0 & 1 & 0 & s_\mathrm{ob}\\
			0 & 0 & 1 & 0\\
			0 & 0 & 0 & 1			
		\end{pmatrix}.
	\end{equation}
	\normalsize
	The coefficients of the polynomial $\tau_3$ are given in Table \ref{tbl::tau3} analogously to Eq.~\eqref{eq::polyGeneral} with coefficients denoted by $b_{ijkl}$.
	
	\begin{table}
		\centering
		\begin{tabular}{ll}
			\toprule
			Coefficients & Values\\[0.5ex]
			\midrule
			$b_{0003}$ & $s_\mathrm{im}^2 \left( a_{0201}-s_\mathrm{im} a_{0300}\right) $\\
			$b_{0021}$ & $s_\mathrm{im}^2 \left( a_{1110}+a_{2001}-s_\mathrm{im} a_{2100}\right) $\\
			$b_{0102}$ & $s_\mathrm{im} \left(3 s_\mathrm{im} a_{0300}-2  a_{0201}\right) $\\
			$b_{0120}$ & $s_\mathrm{im} \left(s_\mathrm{im} a_{2100}- a_{1110}\right) $\\
			$b_{0201}$ & $a_{0201}-3 s_\mathrm{im} a_{0300} $\\
			$b_{0300}$ & $a_{0300} $\\
			$b_{1011}$ & $s_\mathrm{im} \left(2 s_\mathrm{im} a_{2100}- a_{1110}+2 a_{2001}\right) $\\
			$b_{1110}$ & $a_{1110}-2 s_\mathrm{im} a_{2100} $\\
			$b_{2001}$ & $a_{2001}-s_\mathrm{im} a_{2100} $\\
			$b_{2100}$ & $a_{2100}$\\
			\bottomrule
		\end{tabular}
	\caption{Coefficients of $\tau_3$.}
	\label{tbl::tau3}
	\end{table}
	
	The expressions for the coefficients of $\tau_4$ are rather lengthy and not useful for the current discussion, but can be found in Appendix \ref{sec::Appendix} for completeness. We thus have a mapping that describes the fundamental element up to third-order.
	
	\subsection{From Optical Element to Optical System}
	To treat optical systems it suffices to concatenate multiple fundamental elements, keeping in mind the sign conventions described at the end of Section~\ref{sec::fundElemLast}. Each intermediate image plane corresponds to the intermediate object plane of the subsequent mirror. Thus, if one fundamental element is described by Eq.~\eqref{eq::fundElementMod}, then multiple elements are a concatenation of Lie transformations of this form. For example, suppose a two-mirror system where one mirror is described by the generators $\tau_k$ and the other mirror by the generators $\sigma_k$. Then, the map $\mathcal{M}$ of the complete system, up to third-order contributions, reads,
	\begin{equation}
		\mathcal{M}\overset{(3)}{=}\exp([\tau_2,\cdot\,])\exp([\tau_3,\cdot\,])\exp([\tau_4,\cdot\,])\exp([\sigma_2,\cdot\,])\exp([\sigma_3,\cdot\,])\exp([\sigma_4,\cdot\,]).
		\label{eq::twoMirrorEx}
	\end{equation} 
	The coefficients of $\tau_k,\sigma_k$ are completely described by the geometry of the system according to the expressions $b_{ijkl}$.

	Previously, we have stressed the importance of having the Lie transformations in ascending order. This allows us to separate the contributions to the different (ascending) orders of aberrations. The necessary computations to reorder Eq.~\eqref{eq::twoMirrorEx} rely on the procedure for reordering shown in \cite{Barion:22} and make use of the BCH formula \eqref{eq::BCH} and the results of Eq.~\eqref{eq::thrm3}. During these steps, the composition of low-order aberrations into high-order ones follows directly from the application of the BCH formula. 
	
	In more complex optical systems the intermediate image planes for the sagittal and tangential rays need not to be located at the same point along the OAR. As such, the choice of the propagation distances for each fundamental element is seemingly unclear. However, whatever the chosen propagation distance is equal to, the sum of the intermediate image distance of the surface $j$ and the object distance of surface $j+1$ needs always be equal to the total distance between the two surfaces. Since the propagation mappings commute, see \cite{Barion:22,DragtFoundations86,Wolf2004}, it does not matter what distance is chosen for the image propagation of surface $j$ or the object distance for surface $j+1$ as long as their sum remains equal to the distance between the two surfaces. 
	
	\section{Applications}\label{sec::Examples}
	We verify the presented methodology using three examples. We recover the surface expansion coefficients of a spherical ellipsoid for a point-to-point imager and the surface expansion coefficients for a focusing mirror as recently presented in \cite{Caron}. Lastly, we use our proposed method to ray-trace a beam of rays reflected by a biconic mirror and compare with the spot diagram generated using OpticStudio.
	
\begin{figure}[!htb]
		\centering
		\includegraphics[width=0.4\textwidth]{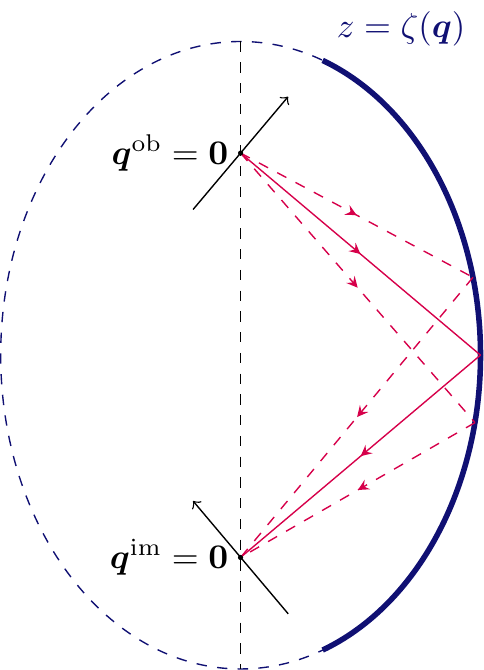}
		\caption{A spherical ellipsoid as perfect imager between its foci. The OAR is in red and the red dashed lines are other rays originating from the object. The object and image planes are represented. The black dashed line is the major axis of the ellipsoid.}
		\label{fig::sphericalEllipse}
	\end{figure}

	The first example will be the problem of perfect point-to-point imaging; see Figure \ref{fig::sphericalEllipse}. Suppose we have an object point on the OAR which is then reflected off a surface onto an image point. A spherical ellipsoid with these two points at its foci will result in perfect imaging \cite{Gomez}, i.e., no aberrations will be present. Therefore, if we choose arbitrarily an object and an image point and impose zero aberrations up to third-order for all rays with initial position $\bm{q}^\mathrm{ob}=\bm{0}$, then the solution for the surface coefficients should be the surface expansion terms up to fourth order of the corresponding spherical ellipsoid.
	
	We fix the object distance, image distance and the surface coefficients $c_{20},c_{02}$ to have the desired paraxial properties. Subsequently, the corresponding map given in Eq.~\eqref{eq::fundElementMod} is applied to the initial coordinates $(\bm{0},\bm{p}^\mathrm{ob})$. The expression for the final position coordinates at the image plane $\bm{q}^\mathrm{im}$ is of the form:
	\begin{equation}
		\label{eq::finalPosEllipsoid}
		\bm{q}^\mathrm{im}=\bm{q}^\mathrm{im}(\bm{0},\bm{p}^\mathrm{ob}).
	\end{equation}
	Eq.~\eqref{eq::finalPosEllipsoid} is a polynomial dependent only on the initial direction $\bm{p}^\mathrm{ob}$ where each monomial coefficient will depend on the chosen parameters and the -- yet undetermined -- higher order coefficients $c_{mn}$ of the reflecting surface; see Eq.~\eqref{eq::surfaceEq}. The requirement of zero aberration, i.e., $\bm{q}^\mathrm{im}=\bm{0}$, simply translates in setting all monomial coefficients in Eq.~\eqref{eq::finalPosEllipsoid} equal to zero. The resulting system of equations will determine the value of the surface expansion coefficients. For example, if we choose a spherical ellipsoid with major axis $a=20$ and minor axis $b=10$, then with the corresponding initial parameters for the system $s_\mathrm{ob}=s_\mathrm{im}=20$ and $c_{20}=-1/20,c_{02}=-1/80$, we get the following system of equations for the unknown coefficients $c_{mn}$ with $m=0,2,4$ and $2\leq m+n\leq 4$:
	\begin{equation}
		\begin{dcases}
			80\, p_x^3 (8000 c_{40}+1)=0,\\
			-80\, p_x p_y^2 \left(2400 \sqrt{3} c_{03}+200 \sqrt{3} c_{21}-16000 c_{22}-1\right)=0,\\
			32000\, p_x p_y c_{21}=0,\\
			80\, p_x^2 p_y \left(2400 \sqrt{3} c_{03}+200 \sqrt{3} c_{21}+16000 c_{22}+1\right)=0,\\
			16000 \,p_x^2 c_{21}=0,\\
			80\, p_y^3 (128000 c_{04}+1)=0,\\
			192000 \,p_y^2 c_{03}=0.
		\end{dcases}
	\label{eq::ellipsoidSystem}
	\end{equation} 
	The solution to Eqs.~\eqref{eq::ellipsoidSystem}, computed with exact arithmetic, gives the surface expansion coefficients shown in Table \ref{tbl::ellipsoidCoeffs}.
	\begin{table}[htb!]
		\centering
		\begin{tabular}{l c} 
			\toprule
			$c_{mn}$ & Values \\ [0.5ex] 
			\midrule
			$c_{2 1}$ & 0 \\
			$c_{0 3}$ & 0 \\
			$c_{4 0}$ & $-1/8000$\\ 
			$c_{2 2}$ & $-1/16000$\\
			$c_{0 4}$ & $-1/128000$\\
			\bottomrule
		\end{tabular}
	\caption{Surface expansion coefficients for the spherical ellipsoid defined according to Eq.~\eqref{eq::surfaceEq}.}
	\label{tbl::ellipsoidCoeffs}
	\end{table}
	The coefficients in Table \ref{tbl::ellipsoidCoeffs} are the same we would get by directly expanding the ellipsoid's equation
	\begin{equation}
		\label{eq::ellipsoidEq}
		z=\zeta(\bm{q})=\frac{b}{a} \sqrt{a^2 - q_y^2 - \frac{a^2}{b^2} q_x^2} - b,
	\end{equation}
 	in terms of $q_x,q_y$ around the origin. In fact, the surface equation \eqref{eq::ellipsoidEq} represents the ellipsoid at the point of impact of the OAR with respect to the coordinate system aligned with its normal at that point; see Figure \ref{fig::sphericalEllipse}.
	
\begin{figure}[!htb]
		\centering
		\includegraphics[width=0.4\textwidth]{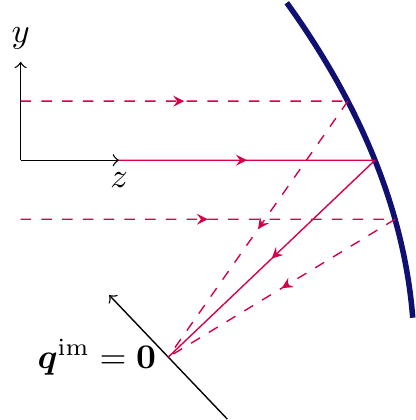}
		\caption{A focusing reflector for an object at infinity as used in \cite{Caron}.}
		\label{fig::parabolicRefl}
	\end{figure}

	The next example reproduces some of the results given in \cite{Caron}. Here, the authors calculate the surface expansion coefficients for a single mirror where again zero third-order aberrations are imposed with the additional condition that the initial momenta are equal to zero, i.e., $\bm{p}^\mathrm{ob}=\bm{0}$; see Figure~\ref{fig::parabolicRefl}. 
	
	The initial surface parameters are the effective radius of curvature $R=-200$, cf. \cite{Caron}, for both the sagittal and tangential planes and the incidence angle of $\theta=-0.2$. The expression for the final position coordinates $\bm{q}^\mathrm{im}$ is now a polynomial in $\bm{q}^\mathrm{ob}$, i.e.,
	\begin{equation}
		\label{eq::finalPosCaron}
		\bm{q}^\mathrm{im}=\bm{q}^\mathrm{im}(\bm{q}^\mathrm{ob},\bm{0}).
	\end{equation}
	Setting all monomial coefficients of Eq.~\eqref{eq::finalPosCaron} to zero, results in the following system of equations for the $c_{mn}$ with $m=0,2,4$ and $2\leq m+n\leq 4$:
	\begin{equation}
		\begin{dcases}
			\frac{q_x^3 \left(-8 R^3 \cos (\theta) c_{40}+\sec ^2(\theta)-1\right)}{2 R^2}=0,\\
			\frac{q_x q_y^2 \sec (\theta) \left(-3 \sec (\theta) \left(2 R^2 (\sin (2 \theta) c_{21}-2 \tan (\theta) c_{03})+\cos (2 \theta)-1\right)-8 R^3 c_{22}\right)}{4 R^2}=0,\\
			-\frac{q_x q_y \left(2 R^2 c_{21}+\tan (\theta)\right)}{R}=0,\\
			-\frac{q_x^2 q_y \sec (\theta) \left(5 \sin (\theta) \left(2 R^2 c_{21}+\tan (\theta)\right)+2 R^2 (3 \tan (\theta) \sec (\theta) c_{03}+2 R c_{22})\right)}{2 R^2}=0,\\
			-\frac{q_x^2 \left(2 R^2 c_{21}+\tan (\theta)\right)}{2 R}=0,\\
			\frac{q_y^3 \sec ^3(\theta) \left(-32 R^2 (2 \sin (\theta) c_{03}+R c_{04})-3 \cos (\theta)+3 \cos (3 \theta)\right)}{8 R^2}=0,\\
			-\frac{3 q_y^2 \left(2 R^2 \sec ^2(\theta) c_{03}+\tan (\theta)\right)}{2 R}=0.
		\end{dcases}
	\label{eq::caronSystem}
	\end{equation}
	The Eqs.~\eqref{eq::caronSystem} are solved using exact arithmetic and give results for the surface expansion coefficients shown in Table \ref{tbl::caronCoeffs}, which agree exactly with those given in \cite{Caron}.
		\begin{table}[htb!]
		\centering
		\begin{tabular}{l c} 
			\toprule
			$c_{mn}$ & Values \\ [0.5ex] 
			\midrule
			$c_{2 1}$ & $2.53388\times 10^{-6}$ \\
			$c_{0 3}$ & $2.43386\times 10^{-6}$ \\
			$c_{4 0}$ & $-6.55111\times 10^{-10}$\\ 
			$c_{2 2}$ & $-3.77553\times 10^{-9}$\\
			$c_{0 4}$ & $-3.02209\times 10^{-9}$\\
			\bottomrule
		\end{tabular}
		\caption{Surface expansion coefficients for the second example defined according to Eq.~\eqref{eq::surfaceEq}.}
		\label{tbl::caronCoeffs}
	\end{table}

\begin{figure}[!htb]
		\centering
		\includegraphics[width=0.4\textwidth]{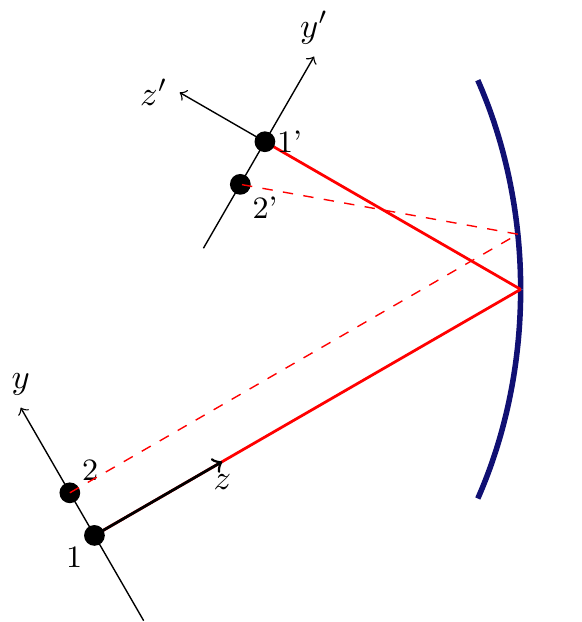}
		\caption{Sketch of the biconic reflector in Eq.~\eqref{eq::biconic}. The point objects $1$ and $2$ are imaged paraxially onto their primed counterparts.}
		\label{fig::biconicRefl}
	\end{figure}

\begin{figure}[htb!]
		\centering
        \includegraphics[width=\textwidth]{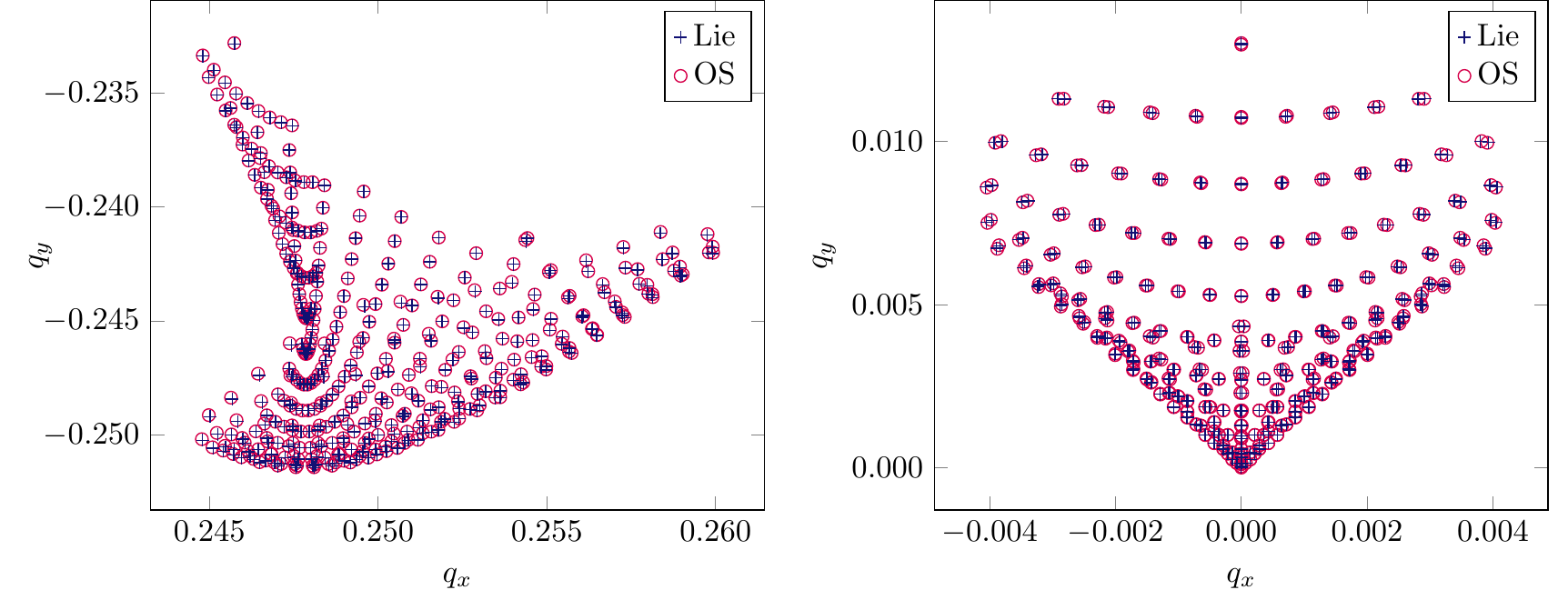}
	\caption{Spot diagrams at image plane. Left: Off-axis object point. Right: On-axis object point.}
	\label{fig::spotDiagrams}
	\end{figure}

	The last example we present in this paper is a comparison between spot diagrams of a biconic mirror when computed using OpticStudio and our Lie method; see Figure~\ref{fig::biconicRefl}. The mapping in Eq.~\eqref{eq::fundElementMod} generates a third-degree polynomial in phase-space variables which can be used as a ray-tracer between object and image plane coordinates. In Figure \ref{fig::spotDiagrams}, we can see the difference in the ray-tracing for an off-axis beam of rays originating from the object point at position $\bm{q}=(-0.5,0.5)$ and direction domain $\bm{p}\in[-0.0075,0.0125]\times[-0.0125,0.0075]$ and for an on-axis beam of rays originating from $\bm{q}=(0,0)$ with direction domain $\bm{p}\in[-0.01,0.01]^2$. For both cases the object and image distances are $s_\mathrm{ob}=200,\,s_\mathrm{im}=100$ and the OAR has an incidence angle equal to $\theta=\pi/6$. The surface equation of the biconic is:
	\begin{equation}
		z=\zeta(\bm{q})=\frac{c_xq_x^2+c_yq_y^2}{1+\sqrt{1-(1+\kappa_x)c_x^2q_x^2-(1+\kappa_y)c_y^2q_y^2}},
		\label{eq::biconic}
	\end{equation}
	with $\kappa_x=\kappa_y=0$ and $c_x=-\dfrac{\sqrt{3}}{200}$, $c_y=-\dfrac{3\sqrt{3}}{800}$. Using the surface expansion coefficients of Eq.~\eqref{eq::biconic} we can determine the necessary coefficients $b_{ijkl}$ for the Lie operators and the resulting spot diagram coincides almost perfectly with the OpticStudio ray tracing. The maximum distance between the coordinates given by the two methods in the examples is $\Delta_{\mathrm{max}}=9\times10^{-5}$.
	
	\section{Conclusions}
	In this paper we extend the procedure presented for rotationally symmetric systems in \cite{Barion:22,DragtFoundations86,Wolf2004} to mirror systems with only planar symmetry. Starting from a set of analytical ray-tracing equations, we expand them up to third-order. The information about these expansions is then encoded into the associated Lie transformations. We derive the generator polynomials for the Lie transformations up to fourth degree. Thus, the method produces third-order analytical expressions for the transverse ray aberrations for an arbitrary mirror with planar symmetry. We calculate coefficients of the generators for a single mirror. These coefficients depend only on the geometrical information of the mirror itself. It is therefore possible to describe an arbitrary optical system as the concatenation of single mirrors since for each mirror the associated generated polynomials are known. Complex phenomena like lower order aberrations combining into higher order ones are captured by the method.
	
	We verified our results with three applications. In the first two, we show how it is possible to use the analytic expressions of the aberrations to determine the freeform coefficients of the mirror surface that eliminate aberrations up to third-order in the case of a point object and an object at infinity. The last example shows that the aberration expressions can also be used for ray tracing (up to the order of accuracy that has been used in the Lie transformations). Here, we see excellent agreement between the Lie-generated spot diagrams and the ones generated by OpticStudio.
	
	The authors now aim to explore the application of the shown method to the limiting case of grazing incidence and to investigate possible applications for the determination of mirror systems free of, or with reduced, third-order aberrations. The latter can serve as advantageous starting designs for complex mirror systems. Additionally, we intend to work out the relation between the Lie aberration coefficients and the wavefront aberration coefficients described in \cite{Moore,MooreErr}.

	\appendix
	\section{Additional Lie Tools}\label{sec::AddLieTools}
	The Baker-Campbell-Hausdorff (BCH) formula describes how two Lie transformations generated by $f$ and $g$ can be combined into a single one generated by $k$:
	\begin{subequations}
		\label{eq::BCH}
		\begin{equation}
			\exp([k,\cdot\,])=\exp([f,\cdot\,])\exp([g,\cdot\,]),
		\end{equation} 
		where
		\begin{equation}
			k=f+g+[f,g]/2+([f,[f,g]]+[g,[g,f]])/12+\cdots\quad.
		\end{equation}
	\end{subequations}
	In the current discussion, through the BCH formula Eq.~\eqref{eq::BCH} the lower degree generators combine into higher degree contributions and therefore the aberrations do so as well \cite{Barion:22,DragtFoundations86,Wolf2004}.
	
	Furthermore, it can be proven that the following identity for a triplet of Lie transformations holds \cite{DragtFinn}:
	\begin{equation}
		\begin{gathered}
			\exp([g,\cdot\,])\exp([f,\cdot\,])\exp(-[g,\cdot\,])=\exp([k,\cdot\,]),\\
			k=\exp([g,\cdot\,])f.
		\end{gathered}
		\label{eq::thrm3}
	\end{equation}
	These tools are important when composing Lie transformations. Consider the example of two mirrors described by the generators $f_2,f_3$ and $g_2,g_3$. To derive the ordered composition map $\mathcal{M}$ of the combined system we proceed as follows:
	\begin{align*}
		\mathcal{M}&=\exp([f_2,\cdot\,])\exp([f_3,\cdot\,])\exp([g_2,\cdot\,])\exp([g_3,\cdot\,])\\
		&=\exp([f_2,\cdot\,])\underbrace{\exp([g_2,\cdot\,])\exp(-[g_2,\cdot\,])}_{=\mathcal{I}}\exp([f_3,\cdot\,])\exp([g_2,\cdot\,])\exp([g_3,\cdot\,])\\
		&=\underbrace{\exp([f_2,\cdot\,])\exp([g_2,\cdot\,])}_{=\exp([k_2,\cdot\,])}\underbrace{\exp(-[g_2,\cdot\,])\exp([f_3,\cdot\,])\exp([g_2,\cdot\,])}_{=\exp([f_3^\mathrm{tr},\cdot\,])}\exp([g_3,\cdot\,])\\
		&=\exp([k_2,\cdot\,])\exp([f^{\mathrm{tr}}_3,\cdot\,])\exp([g_3,\cdot\,])\\
		&\overset{(2)}{=}\exp([k_2,\cdot\,])\exp([k_3,\cdot\,]),\numberthis
  \label{eq::exampleThrm3}
	\end{align*}
	where $f^{\mathrm{tr}}_3=\exp(-[g_2,\cdot\,])f_3$ according to Eq.~\eqref{eq::thrm3} and $k_3=f^{\mathrm{tr}}_3+g_3$ according to Eq.~\eqref{eq::BCH}. The Lie transformations generated by $f_2,g_2$ have related matrices and the product of these matrices determines the generator $k_2$. The insertion of the identity map $\mathcal{I}$ in Eq~\eqref{eq::exampleThrm3} allows us to simultaneously reorder the Lie transformations and apply Eq.~\eqref{eq::thrm3}. Note that, by using the BCH formula there would be additional generators of degree 4 and higher, which can be neglected if we consider ordered compositions of generators up to degree 3; see \cite{Barion:22,DragtFoundations86}.

	\section{Generator Coefficients}\label{sec::Appendix}
	Here the reader will find the non-zero coefficients of $\tau_4$ in Eq.~\eqref{eq::fundElementMod}.
	\clearpage
	
	\begin{landscape} 
		\begin{table}
			\centering
			\begin{tabular}{ll}
				\toprule
				Coefficients & Values\\[0.5ex]
				\midrule
			$b_{0004}$&$  -\frac{1}{8} s_{\mathrm{im}} \left(8 s_{\mathrm{im}}^2 a_{0301}+8 s_{\mathrm{ob}} a_{0200} (s_{\mathrm{im}} a_{0200}+1) (2 s_{\mathrm{im}} a_{0200} (s_{\mathrm{im}} a_{0200}+1)+1)-8 s_{\mathrm{im}} a_{0202}+1\right)+s_{\mathrm{im}}^4 a_{0400}-\frac{s_{\mathrm{ob}}}{8}$\\[0.8ex]
			$b_{0022}$&$  -\frac{1}{4} s_{\mathrm{im}} \left(4 s_{\mathrm{im}}^2 (a_{1210}+a_{2101})+4 s_{\mathrm{ob}} \left(s_{\mathrm{im}} a_{2000}^2+s_{\mathrm{im}} a_{0200}^2 (2 s_{\mathrm{im}} a_{2000}+1)^2+a_{0200} (2 s_{\mathrm{im}} a_{2000}+1)^2+a_{2000}\right)-4 s_{\mathrm{im}} (a_{0220}+a_{2002})+1\right)$\\
			&\qquad$+s_{\mathrm{im}}^4 a_{2200}-\frac{s_{\mathrm{ob}}}{4}$\\[0.8ex]
			$b_{0040}$&$  -\frac{1}{8} s_{\mathrm{im}} \left(8 s_{\mathrm{im}}^2 a_{3010}+8 s_{\mathrm{ob}} a_{2000} (s_{\mathrm{im}} a_{2000}+1) (2 s_{\mathrm{im}} a_{2000} (s_{\mathrm{im}} a_{2000}+1)+1)-8 s_{\mathrm{im}} a_{2020}+1\right)+s_{\mathrm{im}}^4 a_{4000}-\frac{s_{\mathrm{ob}}}{8}$\\[0.8ex]
			$b_{0103}$&$  s_{\mathrm{ob}} a_{0200} (2 s_{\mathrm{im}} a_{0200}+1)^3+s_{\mathrm{im}} (s_{\mathrm{im}} (3 a_{0301}-4 s_{\mathrm{im}} a_{0400})-2 a_{0202})$\\[0.8ex]
			$b_{0121}$&$  s_{\mathrm{ob}} a_{0200} (2 s_{\mathrm{im}} a_{0200}+1) (2 s_{\mathrm{im}} a_{2000}+1)^2+s_{\mathrm{im}} (s_{\mathrm{im}} (-2 s_{\mathrm{im}} a_{2200}+2 a_{1210}+a_{2101})-2 a_{0220})$\\[0.8ex]
			$b_{0202}$&$  6 s_{\mathrm{im}}^2 a_{0400}-3 s_{\mathrm{ob}} a_{0200}^2 (2 s_{\mathrm{im}} a_{0200}+1)^2-3 s_{\mathrm{im}} a_{0301}+a_{0202}$\\[0.8ex]
			$b_{0220}$&$  -s_{\mathrm{ob}} a_{0200}^2 (2 s_{\mathrm{im}} a_{2000}+1)^2+s_{\mathrm{im}} (s_{\mathrm{im}} a_{2200}-a_{1210})+a_{0220}$\\[0.8ex]
			$b_{0301}$&$  4 s_{\mathrm{ob}} a_{0200}^3 (2 s_{\mathrm{im}} a_{0200}+1)-4 s_{\mathrm{im}} a_{0400}+a_{0301}$\\
			$b_{0400}$&$  a_{0400}-2 s_{\mathrm{ob}} a_{0200}^4$\\[0.8ex]
			$b_{1012}$&$  s_{\mathrm{im}} \left(s_{\mathrm{im}}^2 \left(8 s_{\mathrm{ob}} a_{0200}^2 a_{2000}^2-2 a_{2200}\right)+s_{\mathrm{im}} (4 s_{\mathrm{ob}} a_{0200} a_{2000} (a_{0200}+2 a_{2000})+a_{1210}+2 a_{2101})+2 s_{\mathrm{ob}} a_{2000} (2 a_{0200}+a_{2000})-2 a_{2002}\right)+s_{\mathrm{ob}} a_{2000}$\\[0.8ex]
			$b_{1030}$&$  s_{\mathrm{ob}} a_{2000} (2 s_{\mathrm{im}} a_{2000}+1)^3+s_{\mathrm{im}} (s_{\mathrm{im}} (3 a_{3010}-4 s_{\mathrm{im}} a_{4000})-2 a_{2020})$\\[0.8ex]
			$b_{1111}$&$  -2 (s_{\mathrm{im}} (4 s_{\mathrm{ob}} a_{0200} a_{2000} (a_{2000} (2 s_{\mathrm{im}} a_{0200}+1)+a_{0200})-2 s_{\mathrm{im}} a_{2200}+a_{1210}+a_{2101})+2 s_{\mathrm{ob}} a_{0200} a_{2000})$\\[0.8ex]
			$b_{1210}$&$  4 s_{\mathrm{ob}} a_{2000} a_{0200}^2 (2 s_{\mathrm{im}} a_{2000}+1)-2 s_{\mathrm{im}} a_{2200}+a_{1210}$\\[0.8ex]
			$b_{2002}$&$  -s_{\mathrm{ob}} a_{2000}^2 (2 s_{\mathrm{im}} a_{0200}+1)^2+s_{\mathrm{im}} (s_{\mathrm{im}} a_{2200}-a_{2101})+a_{2002}$\\[0.8ex]
			$b_{2020}$&$  6 s_{\mathrm{im}}^2 a_{4000}-3 s_{\mathrm{ob}} a_{2000}^2 (2 s_{\mathrm{im}} a_{2000}+1)^2-3 s_{\mathrm{im}} a_{3010}+a_{2020}$\\[0.8ex]
			$b_{2101}$&$  4 s_{\mathrm{ob}} a_{0200} a_{2000}^2 (2 s_{\mathrm{im}} a_{0200}+1)-2 s_{\mathrm{im}} a_{2200}+a_{2101}$\\[0.8ex]
			$b_{2200}$&$  a_{2200}-4 s_{\mathrm{ob}} a_{0200}^2 a_{2000}^2$\\[0.8ex]
			$b_{3010}$&$  4 s_{\mathrm{ob}} a_{2000}^3 (2 s_{\mathrm{im}} a_{2000}+1)-4 s_{\mathrm{im}} a_{4000}+a_{3010}$\\
			$b_{4000}$&$  a_{4000}-2 s_{\mathrm{ob}} a_{2000}^4$\\[0.8ex]
			\bottomrule
		\end{tabular}
	\end{table}
	\end{landscape}


\begin{backmatter}
\bmsection{Funding} TKI program ``Photolitho M\&CS" (TKI-HTSM 19.0162)

\bmsection{Acknowledgements} The authors thank Teus Tukker (ASML) for his fruitful remarks.

\bmsection{Disclosures} The authors declare no conflicts of interest.

\bmsection{Data availability}Data underlying the results presented in this paper are not publicly available at this time but may be obtained from the authors upon reasonable request.
\end{backmatter}

\bibliography{Bibl}

\end{document}